\documentstyle{mn}

\catcode`\"=\active\let"=\"

\def\3{\ss }

\def\c12{{1\over 2}}

\def\plusplus{\raise 0.3ex\hbox{${\scriptstyle ++}$}{}}

\newcommand{\oversim}[2]{\protect{\mbox{\lower0.5ex\vbox{%
   \baselineskip=0pt\lineskip=0.2ex
   \ialign{$\mathsurround=0pt #1\hfil##\hfil$\crcr#2\crcr\sim\crcr}}}}} 
\newcommand{\simgreat}{\mbox{$\,\mathrel{\mathpalette\oversim>}\,$}} % >~ sign
\newcommand{\simless} {\mbox{$\,\mathrel{\mathpalette\oversim<}\,$}} % <~ sign

\title[Satellite decay in flattened dark matter haloes] 
{Satellite decay in flattened dark matter haloes}%
\author[J. Pe\~{n}arrubia, P. Kroupa \& C.M. Boily]
{Jorge Pe\~{n}arrubia$^1$, Pavel Kroupa$^2$ \& Christian M. Boily$^{1,3}$ \\
$^1$Astronomisches Rechen-Institut, M\"onchhofstrasse 12-14,
   69120 Heidelberg, Germany \\
$^2$Institut f\"ur Theoretische Physik und Astrophysik, Universit\"at Kiel, 
SD-24098 Kiel, Germany \\ 
$^3$Observatoire Astronomique, 11 rue de l'Universit\'e, F-67000 Strasbourg, France}

\begin{document}

\maketitle

\begin{abstract}
We carry out a set of self-consistent $N$-body calculations to compare
the decay rates of satellite dwarf galaxies orbiting a disc galaxy
embedded in a dark matter halo (DMH). We consider both spherical and
oblate axisymmetric DMHs of aspect ratio $q_h=0.6$.  {The
satellites are given} different initial orbital inclinations, orbital
periods and mass. The live flattened DMHs with embedded discs and
bulges are set-up using a new fast algorithm, {\sc MaGalie} (Boily,
Kroupa \& Pe\~{n}arrubia 2001).

We find that the range of survival times of satellites within a
flattened DMH becomes $\sim100\%$ larger than the same satellites
within a spherical DMH. In the oblate DMH, {satellites on polar
orbits} have the longest survival time, whereas {satellites on
coplanar prograde orbits} are destroyed most rapidly. The {orbital
plane of a satellite tilts as a result of anisotropic dynamical
friction,} causing the satellite's orbit to {align with} the plane of
symmetry of the DMH.  {Polar orbits are not subjected to
alignment. Therefore the decay of a satellites in an axisymmetric DMH
may provide a natural explanation for} the observed lack of satellites
within $0-30^\circ$ of their {host} galaxy's disc (Holmberg 1969;
Zaritsky \& Gonz\'alez 1999).

The computations furthermore {indicate} that the evolution of the
orbital eccentricity $e$ is highly dependent of its initial value
$e(t=0)$ and the DMH's shape.  We also discuss some implications of
flattened DMHs for satellite debris streams.

\end{abstract}


\section{Introduction}
Non-spherical mass distributions around galaxies and galaxy clusters
are needed to reconcile the dynamics with lensing statistics (Maller
et al. 2000; Gonz\'alez et al. 1999; Maller et al.  1997; Keeton \&
Kochanek 1998) or galactic disc warps (Binney 1992).  In CDM
cosmogony, aspherical bound dark matter haloes ($\equiv$ DMHs) form as a result
of gravitational clustering.  Dubinsky (1994) finds in his computer
simulations a Gaussian distribution of DMH  {aspect ratios}, $q_h \equiv
c/a >0$, where $c$ and $a$ are the minor and major axes of an oblate
spheroid, of mean $<q_h>=1/2$ and dispersion equal to~0.15.  In a CDM
framework, therefore, DMHs may  {achieve an aspect ratio}
 as high as $q_h =
0.65$, covering a range of values more than adequate to account for
lensing data: for instance, Maller et al. (2000) find that for the
galaxy B$1600+434$, $0.5< q_h <0.75$.  The inferred morphology of the
DMH, however, depends on the details of its radial mass profile and
embedded baryonic galaxy components, such as the disc and bulge of a
spiral galaxy, and is, on the whole, a loosely constrained quantity.
 For instance, Olling \& Merrifield (2000) studying the axis-ratio of the Milky Way by two independent methods (first, by the measurements of its rotational curve and the amount of dark matter in the solar neighborhood and, second, by the variation in thickness of HI emission) find a consistent value of $q_h\sim 0.8$, although it depends strongly on the measurement of the Milky Way's parameters.

In general, different techniques yield a wide range of values for $q_h$.
        Models of decaying neutrinos (Sciama 1990) or cold molecular gas
        (Pfenniger et al. 1994) suggest a minimum value $q_h = 0.2$, whereas
        models of the Milky Way halo  suggest $q_h$ = 0.9 to account for the
        narrow debris stream of the Sagittarius dwarf (Ibata et al. 2001).
        The measure of axis-ratios throughout a DMH is riddled with uncertainties
        due to the unknown profile of the halo. Furthermore, lensing or stellar
        kinematics yield diagnostics that are constrained in the inner region
        mainly, where density gradients are the largest. The Sagittarius
        dwarf data suggest that the inner region of the MW DMH would be
        spherical, yet questions arise as to whether the same hold true on
        scales approaching 100 kpc or more.

The dynamics of galactic satellites may help constrain DMH profiles
on large scales by direct observations of their distribution around the host galaxy.
 Holmberg (1969) and
Zaritsky \& Gonz\'alez (1999) point out that satellites around disc
galaxies are found more often aligned with the poles of the host
galaxy, the so-called 'Holmberg effect'. One possible reason for this
effect is the enhanced satellite-disc coupling for co-planar satellite
orbits.  Quinn \& Goodmann (1986), however, find in their $N$-body
study that discs alone cannot account for the original statistical
distribution of Holmberg's data.

A remedy may be sought in the form of an extended non-spherical DMH.
An anisotropic velocity (and mass) distribution will cause a
satellite's orbit to align with the axes of the velocity ellipsoid of
the host galaxy (Binney 1977). A strategy for exploring the parameter
space of orbits of a population of satellites would be to integrate
the orbit of satellites in a fixed non-spherical potential using
e.g. Binney's treatment for the dynamics of a point source. In the last years, there have been several studies of how dynamical friction influences the orbit and structure of satellites that go in this direction (e.g. Colpi et al 1999 using the theory of Linear Response), 
as well as N-body simulations (e.g. van den Bosch et.al. 1999).  
However
there is at present no body of work for this problem against which to
compare the analytic treatment, as done for spherically symmetric
systems (see for instance Taylor \& Babul 2000).  Furthermore, mass
loss by the satellite is difficult to account for analytically (see
Helmi \& White 1999; Johnston et al. 1999).  In addition, fixed
potentials prevent important feed-backs from the dynamical friction
process. For example, Weinberg (2000) argues that the wake from a
heavy satellite may induce bending modes in the disc.  {Effects
such as} wake, 
disc bending, motion of the primary galaxy's centre of mass,
would in turn influence the satellite's orbit.

In this paper, we study how  {axisymmetric (flattened)} DMHs
 affect the  {orbital} decay and 
survival  of satellites, paying particular attention to the
orbital inclination of the satellite with respect to the disc and DMH
plane of symmetry. We are motivated by the Holmberg effect, and by the
fact that no study of satellite decay exists to date which takes into
account both the velocity anisotropy and the flattened density
structure of the DMH. Given the likely importance of feedback, and the
as yet untested analytical description of dynamical friction in
anisotropic systems, we resort to fully self-consistent calculations
with live multi-component galaxies and satellites. This has become
possible only very recently through the availability of a new
algorithm, {\sc MaGalie}, that allows the construction of large-N
flattened multi-component galaxy models (Boily, Kroupa \&
Pe\~{n}arrubia 2001, hereinafter BKP).

Section~2 introduces the models. In Section~3 we study how flattened
DMHs affect satellite decay, especially in comparison with spherical
DMHs without  velocity anisotropy. We also touch on orbital precession
and its implications for the spread of tidal debris.  
The paper concludes with Section~4.

\section{The models}
A subset of our spherical models are similar to the models of
Vel\'azquez \& White (1999, hereinafter VW) to facilitate an
inter-comparison of different numerical treatments.

\subsection{The primary galaxy model}
In order to minimize computational time when constructing flattened
DMHs with embedded bulges and discs, we apply a new highly-efficient
technique using multi-pole potential expansions to tailor the local
velocity ellipsoid to the required morphology (BKP). The algorithm to add together individual
components in a single galaxy is adapted from Hernquist's method
(Hernquist 1993).  The new code, {\sc MaGalie}, scales linearly with
particle number and hence we can construct flattened DMHs consisting
of $\simgreat10^6$ particles or more, in a short computational time.

For the density distributions of the disc we take
\begin{equation} 
        \rho_d(R,z)=\frac{M_d}{4\pi R_d^2 z_0} \rm{exp}(-R/R_d) 
        \rm {sech}^2(z/z_0),
\end{equation}
$M_d$ being the disc mass, $z_0$ the vertical thickness, and $R_d$ the
exponential scale length in the radial direction. The mass profile
decays exponentially with R and is composed of isothermal sheets along
the vertical direction.  Velocities are assumed to have a Gau{\ss}ian
distribution.  The square of the radial velocity dispersion is taken 
to be proportional to the surface density (see Lewis \& Freeman 1989), $\overline{v_R^2} \propto
\Sigma(R)=\Sigma(0)\rm{exp}(-R/R_d)$, where the constant of
proportionality is determined by fixing Toomre's
Q-parameter at the Solar radius. Following VW, we select
$Q_\odot=Q(R_\odot)=1.5$. The vertical component of the velocity
ellipsoid is $\overline{v_z^2}=\pi G \Sigma(R)z_0$ in agreement with
an isothermal sheet (Spitzer 1942). The azimuthal component is obtained
from the epicyclic approximation $
\sigma_{\phi}^2=\overline{v_R^2}\kappa^2/(4\Omega^2)$ (e.g. Binney \&
Tremaine 1987, hereinafter BT).

For the bulge we adopt the spherical Hernquist profile (Hernquist 1990),
\begin{equation} 
        \rho_b=\frac{M_b}{2\pi} \frac{a}{r(r+a)^3},
\end{equation}  
where $M_b$ is the bulge mass and $a$ the spherical scale length. This
analytical profile fits the de Vaucouleurs law (de Vaucouleurs 1948).
The velocity field is constructed from the Jeans equations by assuming
isotropic Gaussian velocity distributions at each radial distance
(Hernquist 1993).

We use a non-singular isothermal profile for the DMH,
\begin{equation} 
        \rho_h=\frac{M_h \alpha}{2\pi^{3/2}
        r_{\rm cut}}\frac{{\rm exp}(-r^2/r_{\rm cut}^2)}{r^2+\gamma^2},
\label{eqn:rho_h}
\end{equation}
$M_h$ being the DMH mass, $r_{\rm cut}$ the cut-off radius and $\gamma$
the core radius, and
\begin{eqnarray}
        \alpha\equiv\{1-\sqrt{\pi}\beta{\rm exp}(\beta^2)[1-{\rm erf}
                      (\beta)]\}^{-1} =  \\ \nonumber
                  1 + \sqrt{\pi}\beta + (\pi -2) \beta^2 + O(\beta^3) 
\end{eqnarray} 
where $\beta=\gamma/r_{\rm cut}\simless 1/24$  {in our
calculations}. For $\beta = 1/24$ we find 
 $\alpha \simeq 1.076\rightarrow 1$  {already and hence thereafter we 
set $\alpha = 1$ in our analysis.} To construct the flattened
(oblate) DMHs, a non-homologous transformation is applied to
(\ref{eqn:rho_h}) to achieve the desired axis ratio $c/a$ while 
preserving the central density:   
 { 
1) first we  flatten the DMH down the z-axis only, 
 until the desired aspect ratio  $q_h = c/a$ is reached; 2) then 
all axes are stretched by a factor $\lambda$, 
$A = \lambda R$, such that $M/R^3 = M/A^2C
= M/[A^3q_h] = M/[R^3\lambda^3 q_h]. $
 Solving for $\lambda = q_h^{-1/3} \approx 1.18$ when $q_h =
0.6$. The orbital period $\propto 1/\sqrt{G\rho}$, and hence the 
dynamical time, is unchanged in the new equilibrium.  
}
Particle velocities are obtained by
adjusting the initial isotropic distribution (as for the bulge) to the
oblate iso-potential surfaces leading to a stable axisymmetric DMH
with embedded bulge and disc (BKP). 

We use four different isothermal DMH models: a spherical $ $ (G1) and
a flattened DMH (G2) with axis-ratio $q_h = c/a = 0.6$, which lies
within the distribution of flattenings given by CDM models. The third
(G3) and fourth (G4) models have the same properties as G1 and G2,
respectively, but with enlarged cut-off radii.  We can define two
typical distances, the core radius in the  {symmetry plane}
($\gamma_a$) and in the vertical direction ($\gamma_c$). Since
 {concentric iso-density contours} 
have the same axis-ratio throughout the DMH, both core radii are
related by $q_h=\gamma_c/\gamma_a$.
 
\begin{figure}
\vspace{6.8cm}
\includegraphics{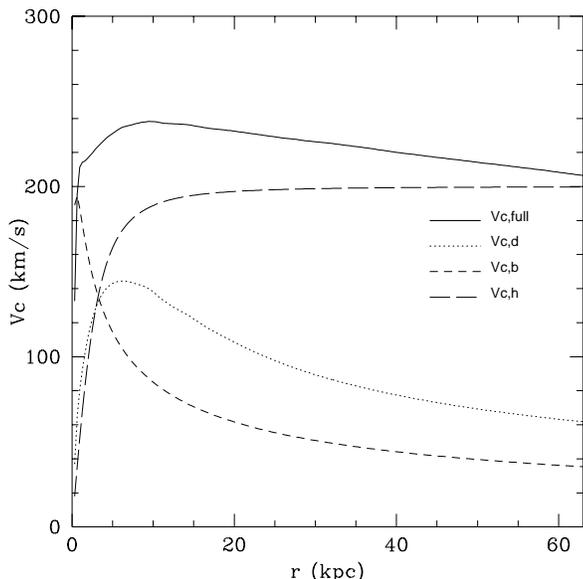}
\caption{Total contribution from the three G1 galaxy components (disc, bulge and halo) to the circular velocity (solid line).  We also plot the 
circular velocity for each galaxy component. On very small scales 
($r <1$ ~kpc) the bulge accounts for the bulk of $V_c$. 
 Further out, the dynamics is dominated by the halo. 
The solar radius is at $R_{\odot}=8.5$ kpc. }
\label{fig:vc}
\end{figure}

Our system of units is such that $M_d=R_d=1$ and $G=1$. According to
Bahcall, Smith \& Soneira (1982), $M_d=5.6\times10^{10}M_\odot$ and
$R_d=3.5$ kpc for the Milky Way which we adopt as a typical primary
galaxy model, so that time and velocity units are, respectively,
$1.3\times10^7$ yr and $262\, \rm{km s^{-1}}$. The half-mass radius of
the disc is located at $R_{0.5} \sim 1.7R_d=5.95 $ kpc, with a
rotation period of 13 time units. Table~1 summarizes the parameters
and Fig.~\ref{fig:vc} plots rotational curves for two models.

As VW point out, there are some caveats to keep in mind concerning the
above models: (i) The DMHs are possibly too small in mass and
extension. Zaritsky \& White (1994) show, by studying satellite orbits
in the Local Group and external galaxies, that DMH limits may extend
beyond 200 kpc with masses over $2 \times 10^{12}
\rm{M}_\odot$. However, as VW comment, the velocity curves of our
DMHs G1~and~G2 are consistent with the largest velocities observed
for stars in the solar neighbourhood (Carney \& Lathman 1987),
and they are possibly massive enough to give realistic velocities of
satellites on eccentric orbits.  (ii) The DMHs may be too
concentrated. Persic, Salucci \& Stel (1996) argue for a DMH core
radius of $\gamma=(1 \to 2) \times R_{\rm{opt}}$, $R_{\rm{opt}}=3.2\,R_d$
, where $R_d$ is the disc scale-length.  However the DMH parameters
were selected to avoid bar formation in the disc. We observed that a
less concentrated DMH or bulge allows a stable disc to form a bar
after few satellite passages. With our $\gamma$, the presence of a bar
is avoided at least until the destruction of the satellite.
\\

\begin{table}
\begin{tabular}{||l |l |l |l ||} \hline \hline
        & Symbol & Value(ph.u) & Value (m.u) \\ \hline  
Disc    & $N_d$ & 100000 & \\
        & $M_d$ & $5.60 \times 10^{10} \rm{M}_\odot$ & 1.00 \\
        & $R_d$ & 3.50 kpc & 1.00\\
        & $z_0$ & 1.40 kpc & 0.40 \\
        & $Q_\odot$ & 1.50 & 1.50 \\      
        & $R_\odot$ & 8.50 kpc & 2.43 \\ \hline
Bulge   & $N_b$ & 33328 & \\
        & $M_b$ & $1.87\times 10^{10} \rm{M}_\odot$ & 1/3 \\
        & $a$ & 0.53 kpc & 0.15 \\ \hline
DMH (G1)   & $N_h$ & 1400000 & \\
(spherical)\\
        & $M_h$ & $7.84 \times 10^{11} \rm{M}_\odot$ & 14.00 \\
        & $\gamma$ & 3.50 kpc & 1.00 \\
        & $q_h$ & 1.00 & 1.00 \\
        & $r_{\rm{cut}}$ & 84.00 kpc & 24.00 \\ \hline
DMH (G2) & $N_h$ & 1400000 & \\
(oblate)\\
        & $M_h$ & $7.84 \times 10^{11} \rm{M}_\odot $ & 14.00 \\
        & $q_h$ & 0.60 & 0.60 \\
        & $\gamma_a$  & 3.80 kpc & 1.10\\
        & $\gamma_c$  & 2.28 kpc & 0.65\\ 
        & $r_{\rm{cut}}$ & 84.00 kpc & 24.00 \\ \hline
DMH (G3) & $N_h$ & 1400000 & \\
(spherical)\\
        & $M_h$ & $7.84 \times 10^{11} \rm{M}_\odot$ & 14.00 \\
        & $\gamma$ & 3.50 kpc & 1.00 \\
        & $q_h$ & 1.00 & 1.00 \\
        & $r_{\rm{cut}}$ & 133.00 kpc & 38.00 \\ \hline
DMH (G4) & $N_h$ & 1400000 & \\
(oblate)\\
        & $M_h$ & $7.84 \times 10^{11} \rm{M}_\odot $ & 14.00 \\
        & $q_h$ & 0.60 & 0.60 \\
        & $\gamma_a$  & 3.80 kpc & 1.10\\
        & $\gamma_c$  & 2.28 kpc & 0.65\\ 
        & $r_{\rm{cut}}$ & 133.00 kpc & 38.00 \\ 
\hline \hline
\newline
\newline
\end{tabular}
\caption{Primary galaxy models.  {Oblate models have an aspect
ratio $q_h = 0.6$. The units are such that } 
  {Ph.u.} means 'physical units', and  {m.u.} 'model units'.} 
\label{tab:galmods}
\end{table}

\subsection{Satellite models}
We use self-consistent King models (King 1966) to represent our dwarf
galaxies. 
These models fit early-type dwarf galaxies (Binggeli et
al. 1984), where $r_c$ and $r_t$ are the core and tidal radii,
respectively.  For a comparison with the work of VW we adopt $c=0.8$.

\begin{figure}
\vspace{7.5cm}
\includegraphics{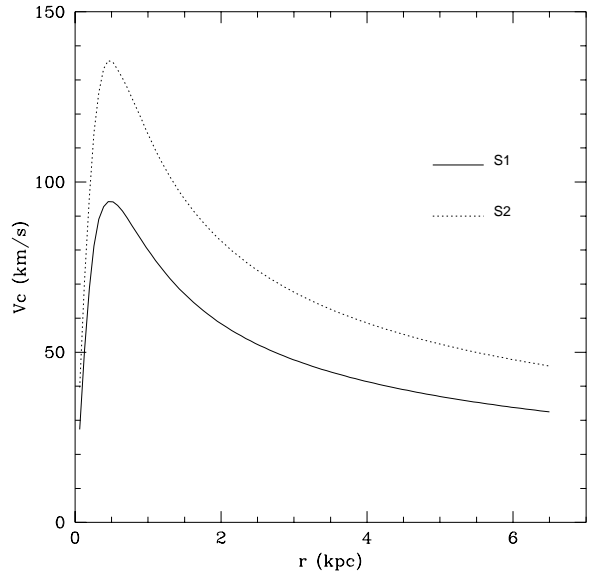}
\caption{Rotational curve of the satellite models S1 and S2 (see Table
2 for the characteristics of each one).}
\label{fig:vcsat} 
\end{figure}

To construct the models we choose the satellite mass $M_s$, $r_c$,
$r_t$ and thus $c$.  The tidal radius is determined by computing the
density contrast, $\rho_s(r_t)/\overline{\rho_g}(r_a) \sim 3$, at the
apo-centric distance ($r_a=55$~kpc) at $t=0$, $\overline{\rho_g}(r)$
being the averaged density of the galaxy (same procedure as VW). This
guarantees that all satellite particles are bound at $t=0$. Tables for
the numerical rendition of the corresponding King profiles can found
in BT or in the original paper of King (1966).  Table~2 summarizes the
parameters, while Fig.~\ref{fig:vcsat} plots rotational curves. Note
that we use the same $M_s$, $r_c$ and ``$r_t$'' despite placing the
satellites at different apogalactica $r_a\ge 55$~kpc
(Section~\ref{sec:orbits}), which increases the true tidal radius of
the satellite, though the stability condition at $t=0$ is still well-accomplished. We do this rather than using different $r_c$ or $r_t$
in order to study the same satellites on different orbits.

Our satellites are much more massive than the Milky Way dSph
satellites which have $M_s\simless 10^8\,M_\odot$, but our adopted
values are typical for the satellites that enter distant samples such
as used by Holmberg (1969) and Zaritsky \& Gonz\'alez (1999).

\begin{table}
\begin{tabular}{||l |l |l |l ||} \hline \hline
  &Symbol & Value(ph.u) & Value (m.u) \\ \hline 
  S1 &  $N_s$ & $40000$ & \\
     &$M_s$ & $5.60\times 10^9 \rm{M}_{\odot}$ & 0.10 \\ 
     &$\Psi(0)/\sigma_0^2$ & 5.00  & 5.00 \\
     &$r_c$ & 1.00 kpc & 0.29 \\ 
     &$r_t$ & 6.31 kpc & 1.80 \\
     &$c$ & 0.80 & 0.80 \\ 
     &$<r>$ & 1.64 kpc & 0.47 \\     
     &$\sigma_0$ & $52.00 \rm{km s^{-1}}$ & 0.20 \\ \hline
  S2 &$N_s$ & $40000$ & \\
     &$M_s$ & $1.12\times 10^{10} \rm{M}_{\odot}$ & 0.20 \\ 
     &$\Psi(0)/\sigma_0^2$ & 5.00  & 5.00 \\
     &$r_c$ & 1.00 kpc & 0.29 \\ 
     &$r_t$ & 6.31 kpc & 1.80 \\
     &$c$ & 0.80 & 0.80 \\ 
     &$<r>$ & 1.64 kpc & 0.47 \\     
     &$\sigma_0$ & $74.00 \rm{km s^{-1}}$ & 0.28 \\
\hline \hline
\newline
\newline
\end{tabular}
\caption{Satellite models. $\Psi(0)=\Phi(r_t)-\Phi(0)$, $\Phi(0)$
being the central potential and $\Phi(r_t)$ the potential at the tidal
radius (following BT notation); $\sigma_0$ is the velocity dispersion
at the centre, and $<r>$ the average radius of the satellite.}
\end{table}

\subsection{Numerical method and orbital parameters}
\label{sec:orbits}
We use {\sc Superbox} (Fellhauer et al. 2000) to evolve the
galaxy-satellite system. {\sc Superbox} is a highly efficient particle
mesh-code based on a leap-frog scheme, and  {has been} already implemented in
an  {ex}tensive study of satellite disruption by Kroupa (1997) and Klessen
\& Kroupa (1998).

Our integration time step is $0.39$ Myr which is about $1/25$th the
dynamical time of satellite~S2.  We have three resolution zones, each
with $64^3$ grid-cells: (i) The inner grid covers out to 3~radial disc
scale-lengths, which contains $\approx 90~\%$ of the disc mass,
providing a resolution of 350~pc per grid-cell. (ii) The middle grid
covers the whole galaxy, with an extension of 24 disc scale-lengths
(84~kpc) for the models G1 and G2, giving a resolution of 2.8~kpc per
grid-cell. The satellite always orbits within this grid except when it
reaches the disc, avoiding cross-border effects (see Fellhauer et
al. 2000). For the models G3 and G4, the middle grid extends to
141~kpc and has a resolution of 4.7~kpc per grid-cell. The orbits of
the satellites are located within this zone.  (iii) The outermost grid
extends to 348~kpc and contains the local universe, at a resolution of 11.6 Kpc.

As for the satellite grid-structure, the resolutions are 816~pc per
grid-cell for the inner grid that extends to 24.48~kpc, 1.2~kpc per
grid-cell for the middle grid which extends to 36~kpc, and 11.6~kpc
per grid-cell for the outermost grid that covers the local
universe. Only the inner and middle grids move along with the
satellites, remaining positioned on their centre-of-density locations.
The outer grid is identical for primary galaxy and satellite.  

Klessen \& Kroupa (1998)
        compared calculations performed with SUPERBOX with direct-integration
        N-body calculations and found good agreement. Specifically, they
        verified that varying the grid resolution by factors of a few 
did not lead to unstable satellite models.
        The stability of the satellite models does not depend strongly on the
        values adopted here.   Furthermore, based on the comparison with the
        direct-integration method, the
        heating introduced by two-body effects prove entirely negligible for
        the  model satellites we consider.
The selection of grid parameters ensures the conservation of energy
and angular momentum for satellites in isolation over times as long as our
calculations to a high degree.
Conservation of total energy and angular momentum is better than 1\%
        for all the models.

 The disc is  {poorly} resolved in the z--direction and we
do not study its evolution in any detail.  
  {We verified that the disc parameters do not evolve for 
galaxies in isolation (no satellites). Since {\sc Superbox} is a mesh code, a poor $z$-resolution for the disc is expected due to the limited number of grids. This provokes the disc modeled here to be unrealistically thick, however it does provide a 
quadrupolar (non-spherical) potential of the appropriate
magnitude. A mesh code has the advantage that it does not 
introduce self-heating since it does not calculate two-body interactions, which would have been significant in the disc given the 
finite number of particles used (see the discussion in VW).} The effects on the satellite dynamics due to two-body interactions are drastically reduced by the low mass of the halo particles (see Steinmetz \& White 1997). Furthermore the disc heating by halo particles is minimized since each component particle masses are in a one-to-one ratio.  

The decay of satellites with various masses through dynamical friction
in an extended spherical DMH is studied by Fellhauer et
al. (2000), who
found good agreement between SUPERBOX calculations and Chandrasekhar's (1960)
formula for dynamical friction when the Coulomb logarithm $\Lambda$ is set
to $\ln\lambda = 1.6$ (cf. eq ~\ref{eqn:dfric} below). This agrees with the findings by
VW for similar calculations.

We carry out a set of calculations varying the parameters of the
satellite and the primary galaxy that influence the satellite--primary
galaxy interaction. These parameters are: (i) the initial orbital
inclination ($\theta_i$), defined as the angle between the initial
angular momentum vector of the satellite and the initial angular
momentum of the disc, (ii) the satellite's mass, (iii) the satellite's
apo-galactic distance, (iv) its orbital eccentricity, and (v) the
DMHs ellipticity, $1-q_h$.

Before injecting the satellite into the primary galaxy we allow the
galaxy and satellite to settle into a stationary state by integrating
the isolated systems for a few dynamical times with {\sc Superbox} (as
in Kroupa 1997). Examples of the stationarity of multi-component
galaxies are given in BKP. The
satellite is then placed at apo-galacticon with a velocity as
described next.

The orbit of the satellites are rosettes. VW define the `circularity'
of the orbit as $\epsilon_J \equiv J/J_C(E)$, $J$ being the
satellite's angular momentum and $J_C$ the corresponding angular
momentum for a circular orbit with the same energy $E$ as the
satellite's orbit.  In practice, we take the circular velocity from
the rotational curve plotted in Fig.~\ref{fig:vc} at the satellite's
initial distance, and multiply it by $\epsilon_J$. This procedure
gives an eccentric orbit with the same energy.  
The  {parameters of the} numerical experiments are listed in
Table~\ref{tab:numexp}.

\begin{table}
\begin{tabular}{||l |l |l |l |l |l |r ||} \hline \hline
Name & Gal.       & Sat.       &$\theta_i$ &$e$ &$r_p$ &$r_a$ \\ 
     & model      & model      &           &    &[kpc] &[kpc] \\  \hline
G1S100 & G1 & S1 & $0^{\circ}$ & 0.7        &17  &55 \\
G1S1180 & G1 & S1 & $180^{\circ}$ & 0.7     &17  &55 \\
G1S145 &  G1 & S1 & $45^{\circ}$  & 0.7     &17  &55 \\               
G1S1135 & G1 & S1 & $135^{\circ}$ & 0.7     &17  &55 \\
G1S190 & G1 & S1 & $90^{\circ}$   & 0.7     &17  &55 \\ \hline
G2S100 & G2 & S1 & $0^{\circ}$  & 0.7       &17  &55 \\
G2S115 & G2 & S1 & $15^{\circ}$  & 0.7      &17  &55 \\
G2S130 & G2 & S1 & $30^{\circ}$  & 0.7      &17  &55 \\ 
G2S145 & G2 & S1 & $45^{\circ}$   & 0.7     &17  &55 \\
G2S160 & G2 & S1 & $60^{\circ}$   & 0.7     &17  &55 \\
G2S190 & G2 & S1 & $90^{\circ}$   & 0.7     &17  &55 \\ 
G2S1135 & G2 & S1 & $135^{\circ}$   & 0.7     &17  &55 \\ \hline
G1S100e & G1 & S1 & $0^{\circ}$  & 0.45     &30  &55 \\
G1S190e & G1 & S1 & $90^{\circ}$ & 0.45     &30  &55 \\ \hline
G2S100e & G2 & S1 & $0^{\circ}$  & 0.45     &30  &55 \\
G2S190e & G2 & S1 & $90^{\circ}$ & 0.45     &30  &55 \\ \hline
G1S100c & G1 & S1 & $0^{\circ}$  & 0        &55  &55 \\
G1S145c & G1 & S1 & $45^{\circ}$ & 0        &55  &55 \\
G1S190c & G1 & S1 & $90^{\circ}$ & 0        &55  &55 \\ \hline
G2S100c & G2 & S1 & $0^{\circ}$  & 0        &55  &55 \\
G2S190c & G2 & S1 & $90^{\circ}$ & 0        &55  &55 \\ \hline
G1S200 & G1 & S2 & $0^{\circ}$  & 0.7       &17  &55 \\ 
G1S245 & G1 & S2 & $45^{\circ}$  & 0.7      &17  &55 \\
G1S290 & G1 & S2 & $90^{\circ}$ & 0.7       &17  &55 \\ \hline
G2S200 & G2 & S2 & $0^{\circ}$  & 0.7       &17  &55 \\ 
G2S245 & G2 & S2 & $45^{\circ}$  & 0.7      &17  &55 \\ 
G2S290 & G2 & S2 & $90^{\circ}$  & 0.7      &17  &55 \\ \hline
G3S200 & G3 & S2 & $0^{\circ}$  & 0.8       &20  &110\\
G3S245 & G3 & S2 & $45^{\circ}$ & 0.8       &20  &110\\  
G3S290 & G3 & S2 & $90^{\circ}$  & 0.8      &20  &110\\ \hline
G4S200 & G4 & S2 & $0^{\circ}$  & 0.8       &20  &110\\
G4S245 & G4 & S2 & $45^{\circ}$  & 0.8      &20  &110\\
G4S290 & G4 & S2 & $90^{\circ}$  & 0.8      &20  &110\\ 
\hline \hline
\newline
\newline
\end{tabular}
\caption{The numerical experiments. The peri- and apo-galactica are    
$r_p$ and $r_a$, respectively, and $e=1-r_p/r_a$ is the
orbital  {ellipticity  (BT, p.21)}.} 
\label{tab:numexp}
\end{table}

\section{Satellite Decay} 
 {We discuss our results in general terms below before going into
detailed consideration of the mass loss and survival of satellites 
(Section 3.2),
and the orbital evolution of the inclination angle, 
eccentricity and precession, respectively (Sections 3.3 to 3.5). }
  Section~\ref{sec:str} takes a brief look
at the implications for tidal streams of dissolving satellites.

\subsection{Introductory comments}
\label{sec:intrcom}
{We denote by `G1S145' the compound primary galaxy made, in this case,
of a spherical DMH plus embedded disc and bulge, G1, and satellite S1,
in an orbital plane initially set at an inclination angle $\theta =
45^\circ$ with respect to the plane of symmetry of the system.  In
what follows we take this model as reference, but all models followed
a similar evolution.

There are two main physical mechanisms that regulate the satellite's
orbital decay: (i) dynamical friction from the disc, bulge and 
DMH, and (ii) tidal interactions, causing internal heating and mass
loss. 
The evolution of the satellite's orbital radius and mass profile 
 highlight the  basic characteristics of these two processes.  
  Dynamical friction causes a steady decrease of the satellite's 
apo- and peri-centres in time as shown  on Fig.3 (dotted line). 
 (Lengths are given in model units on the figure but the time is in
Gyr.) 
From $t=0$ and until  $t < 2 $Gyr, both quantities, apo- and peri-centres,
 decrease monotonically. When  $t > 2$ Gyr, the orbital radius $r \approx 5$ or smaller, and the orbital
decay is not monotonic anymore. 
The proximity to 
 the disc means that non-radial forces affect strongly the remaining 
 evolution, along with  the structure of the satellite. 

To measure changes in the structure of the satellite, we
plotted the ten-percentile Lagrange radii centred on the density
maximum of the satellite (Fig.3, solid lines). 
At $t \approx 3$ Gyr, the galactic tidal field has inflated the satellite to the
 extent that half of its initial mass is spread throughout the volume 
 circumscribed by its orbit.  We note that the inner 10 per cent
Lagrange radius is largely unaffected until the very late stages of 
 integration. Our strategy for determining the orbital parameters of 
the satellite therefore 
consisted in locating the position of the density maximum of the
inner-most Lagrange radius, which then defines a reference
coordinate. }

\begin{figure}
\vspace{9cm}
\includegraphics{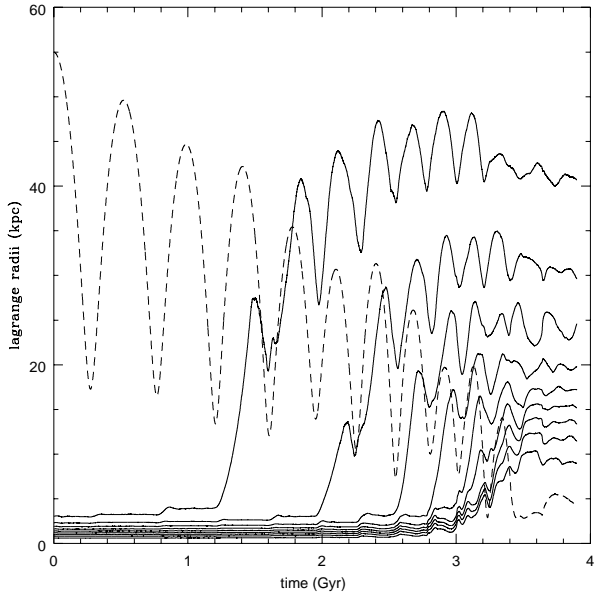}
\caption{Evolution of the satellite's Lagrange radii (solid curves,
defined as the radius at which the spherically enclosed mass amounts
to 10\%, 20\%....,90\%) for the model G1S145. The dotted line
represents the distance of the satellite's centre of density to centre
of the primary galaxy.  Distances are in model units. The overall
evolution is similar in all other models (Table~\ref{tab:numexp}).}
\label{fig:lagr}
\end{figure}

\subsection{Mass loss and disruption times}
\label{sec:mloss}
To calculate the mass remaining bound to the satellite, $M_s(t)$, we
compute the potential energy $\Phi_i < 0 $ 
of each satellite particle presumed bound to
the satellite, and its kinetic energy ($T_i$) in the
satellite frame. Following VW, particles with
$E_i=T_i+m_s(\Phi_i+\Phi_{{\rm ext}})>0$ are labeled unbound, 
where $m_s$ is the mass of
one satellite particle. Particles with $E_i >0$ are removed 
and the procedure repeated until only negative energy particles are
left.  $\Phi_{{\rm ext}} = GM_g(r<r_s)/r_s > 0 $  is the
external potential from the  {primary} galaxy at the satellite's
centre-of-density ($r_s$). 
All the particles of the satellite are thus assumed to
feel the same external potential, which is a useful and sufficiently
accurate approximation, taking into account that most of the bound
particles are located very close to this point. 
For example, in
Fig.~\ref{fig:lagr} most of the satellite's mass lies at a distance
less than 4 kpc from the position of the centre-of-density until the
satellite's disruption. This approximation fails whenever the
satellite's size is comparable to its distance to the galaxy centre.

Satellites lose mass due to the galaxy's tidal forces. The mass loss
happens mostly at perigalacticon, since the gradient of the galaxy's
gravitational force reaches a maximum at that point (see
Fig.~\ref{fig:lagr}).  {This is seen indirectly in the oscillations 
of Lagrange radii, always in phase with the orbit of the satellite:  
the satellite fills its Roche lobe and consequently 
responds strongly to the changing tidal field.} Thus 
a decrease of the apo-galacticon distance
implies an enhanced mass loss.  
The evolution of satellites exposed to strongly varying
tidal fields is discussed at length by Piatek \& Pryor (1995) for one
perigalactic passage, whereas long-term satellite harassment is
addressed by Kroupa (1997) and Klessen \& Kroupa (1998). Consequently,
we  {will}
 not study the internal evolution of the satellites  {apart from 
the bound mass fraction}.

\begin{figure}
\vspace{9.0cm}
\includegraphics{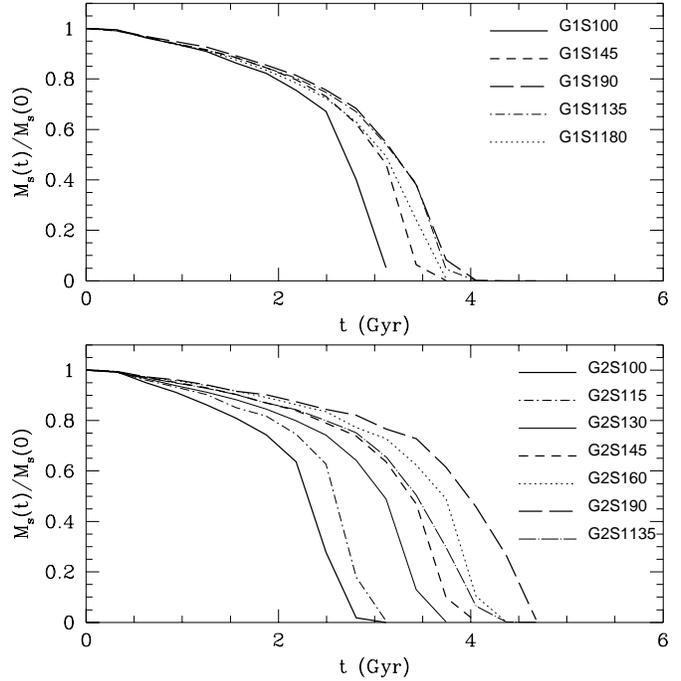}
\caption{{\bf a:} Evolution of the satellite mass for $M_s=0.1M_d$ and
eccentricity $e \simeq 0.7$. }
\label{fig:ms1}
\end{figure}

\begin{figure}
\vspace{4.5cm} \includegraphics{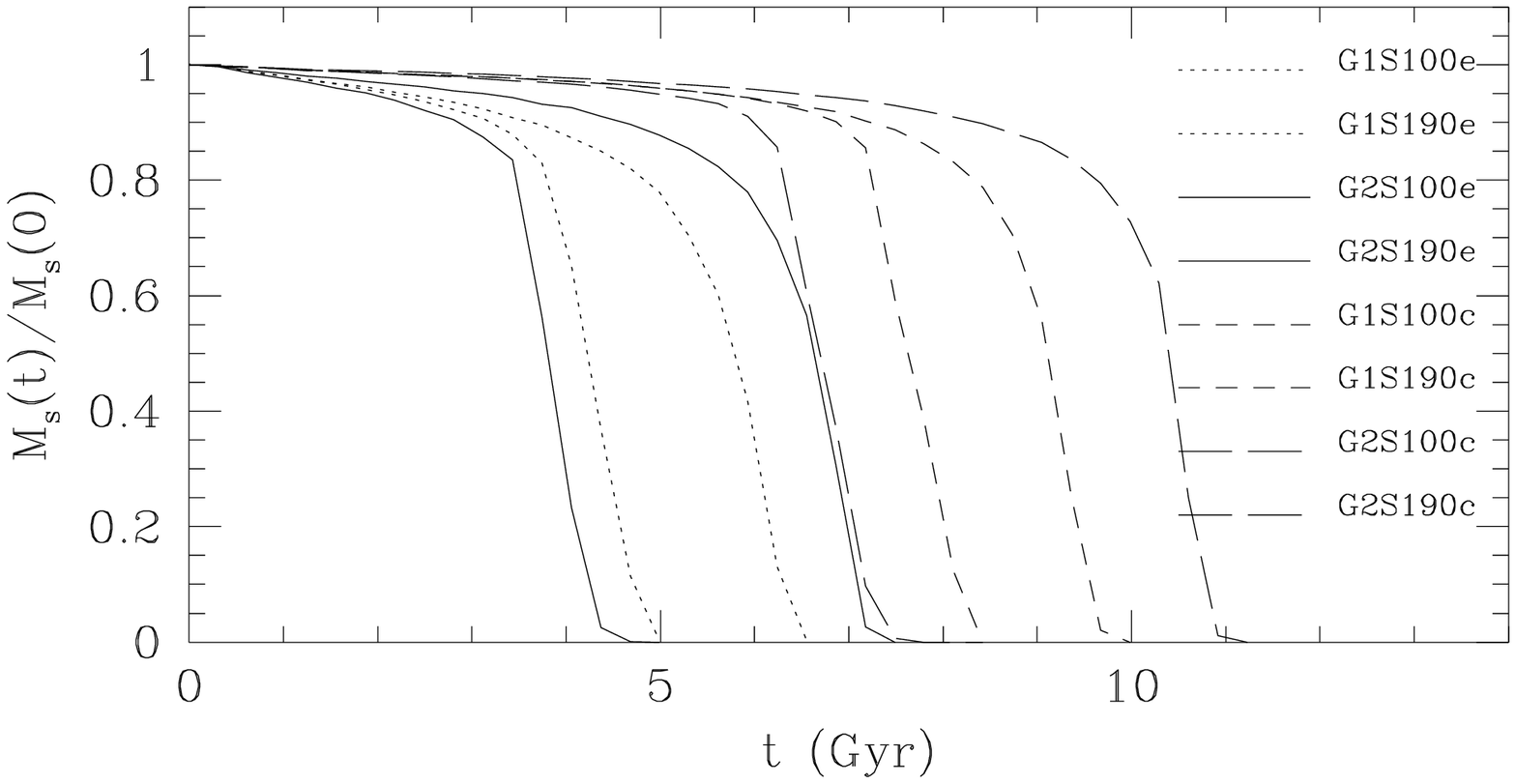} \contcaption{{\bf b:} As Fig.~\ref{fig:ms1}a
for satellites with $M_s=0.1\,M_d$ and initial eccentricity $e \simeq
0.45$ and $e=0$.  (Note that the time-axis has changed scale.)}
\end{figure}

\begin{figure}
\vspace{9.0cm} \includegraphics{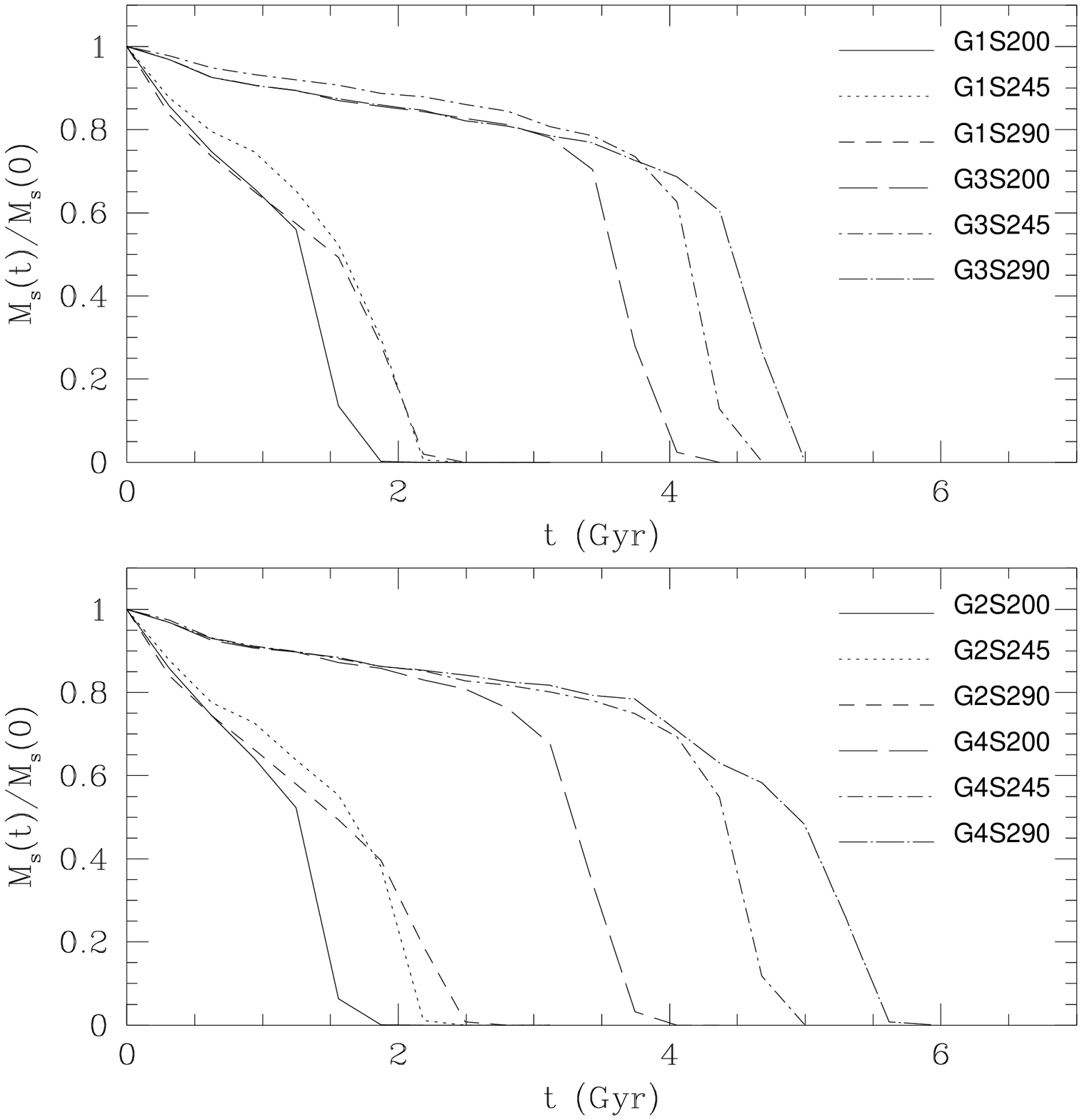} \contcaption{{\bf c:} As Fig.~\ref{fig:ms1}a for
satellites with $M_s=0.2M_d$. (Note that the time-axis has changed
scale.)}  
\end{figure}

\subsubsection{Satellites with $M_s=0.1\,M_d$}
Fig.~\ref{fig:ms1} shows the evolution of the satellite mass for
different initial orbital inclinations for satellites with $M_s\equiv
M_s(0) =0.1\,M_d$ and eccentricity $e \simeq 0.7$. From this figure we
can assert that: (i) The satellites are disrupted completely at about
the same time they reach the galactic disc (Fig.~\ref{fig:lagr}).
(ii) For all the models, the survival time is, at least, 1~Gyr
($25\%$) longer than the equivalent simulations of VW (upper panel of
Fig.~\ref{fig:ms1}a). We consider this difference to be indicative of
the uncertainty intrinsic to methods that approximate collisionless
dynamics.  The difference comes about, in part, due to different
numbers of particles, but also due to the  {spatial} resolution of the
method.  Prugniel \& Combes (1992) and Whade \& Donner (1996) find
that dynamical friction is artificially increased due to numerical
noise if the particle number is small. Similar differences were also
noted in the computations by Klessen \& Kroupa (1998) of satellite
harassment using different codes. However, we observe that the range
of disruption times for our models G1S1 (as used by VW) is
approximately the same, indicating that disc effects are well
reproduced by our code and giving confidence to the following results
we obtain using flattened DMHs. (iii) Flattened DMHs spread the
range of disruption times. In Fig.~\ref{fig:ms1}a we can see that, for
satellites with $M_s=0.1\,M_d$ embedded within spherical DMHs this
range is $\sim 0.9$ Gyr (upper panel), polar satellites having the
longest survival time. For satellites with the same mass but within
flattened DMHs the range grows to $\sim 1.9$ Gyr (lower panel).
(iv) Satellites with a high orbital inclination within flattened
DMHs have longer survival times than satellites within spherical
DMHs with the same initial orbit. For instance, taking the polar
satellite as the extreme case, G2S190 survives $\sim 0.6-1$ Gyr longer
than G1S190.  (v) Satellites with low orbital inclination suffer the
contrary effect: those within spherical DMHs survive longer than
those within flattened DMHs. Taking the prograde and coplanar orbit
as the extreme case, G1S100 survives $\sim 0.4$ Gyr longer than
G2S100.

In Fig.~\ref{fig:ms1}b we compare polar and coplanar satellites within
flattened and spherical DMHs with orbital eccentricity $e\approx
0.45$ and~0 to obtain an indication of the dependency of the life-time
on $e$ (orbits with intermediate inclination also have intermediate
survival times, Fig.~\ref{fig:ms1}a). As expected, less eccentric
orbits lead to longer survival times, since the perigalactic distance
is larger and, moreover, tidal forces are weaker.  Furthermore, the
survival times show a larger spread.  Less eccentric orbits survive
longer, so that anisotropic dynamical friction has a longer time to
act. We can see that coplanar satellites within a spherical DMH (model
G1S100e) survive $\sim 0.6$ Gyr longer than a coplanar satellite
within a flattened DMH (model G2S100e), while the survival time of a
polar satellite within a spherical DMH (model G1S190e) is $\approx
1$~Gyr shorter than the corresponding satellite in the flattened DMH
(model G2S190e). Thus, the range of survival times increases from
about 1.5~Gyr to 3~Gyr. This range becomes even larger for circular
orbits.

\begin{figure}
\vspace{8.5cm} \includegraphics{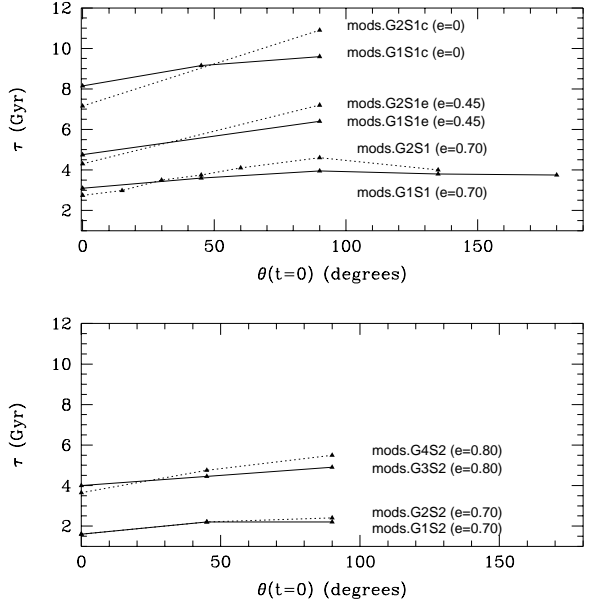}
\caption{The time $\tau$ when the satellite mass reaches 10~per cent
of its initial value, $M_s(\tau)=0.1\,M_s$, is plotted vs the initial
orbital inclination. Upper panel is for satellite models S1 in primary
galaxies G1 and G2, whereas the lower panel shows the results for
satellites S2. Note that in all cases $\tau$ increases with increasing
$\theta<90^{\rm o}$ for galaxies embedded in a spherical and a
flattened DMH, due to dynamical friction on the disc. The effect of
this is particularly nicely seen from the different slopes,
$d\tau/d\theta$, for prograde ($\theta=0-90^\circ$) and retrograde
($\theta=90-180^\circ$) orbits.  The increase is significantly larger
for satellites orbiting in flattened DMHs, and becomes larger for
decreasing orbital eccentricity (Table~\ref{tab:dectemps}) and
decreasing satellite mass, which allows longer coupling of the
satellite to the anisotropic velocity field in the DMH.}  

\label{fig:tvstheta}
\end{figure}

This state of affairs is summarized in Fig.~\ref{fig:tvstheta} for all
satellite models, whereas Table~\ref{tab:dectemps} compares the decay
times for~S1 satellites in dependence of the orbital eccentricity and
inclination. The table nicely shows that the survival time increases
significantly with decreasing eccentricity. It also shows that oblate
DMHs lead to consistently larger differences, $\Delta\tau$, between
the decay times for polar and coplanar orbits, $\Delta\tau$
consistently being approximately 100~per~cent larger in flattened DMHs
than in spherical DMHs ($\Delta\tau_{\rm obl}\approx
2\,\Delta\tau_{\rm sph}$).  This is the key result of this study.

\begin{table}
\begin{tabular}{||l |l |r |r |r ||} 
\hline \hline
model  &$e(t=0)$  &$\tau_0\equiv$    &$\tau_{90}\equiv$      &$\Delta\tau\equiv$\\
       &          &$\tau(\theta=0)$  &$\tau(\theta=90)$      &$\tau_{90}-\tau_0$\\
       &          &[Gyr]             &[Gyr]                  &[Gyr]\\
\hline                                                       
G2S1c(obl)  &0         &7.0               &11.0                   &4.0\\
G1S1c(sph)  &0         &8.0               &9.6                    &1.6\\
\\                                                                
G2S1e(obl)  &0.45      &4.3               &7.3                    &3.0\\
G1S1e(sph)  &0.45      &4.8               &6.5                    &1.7\\
\\                                                                
G2S1(obl)  &0.7        &2.7               &4.6                    &1.9\\
G1S1(sph)  &0.7        &3.1               &4.0                    &0.9\\
\hline \hline
\newline
\newline
\end{tabular}
\caption{Summary of decay times for satellite models~S1
($M_s=0.1\,M_d$) in oblate (obl) and spherical (sph) DMHs with
different initial orbital eccentricity $e$ and orbital inclination
$\theta$.  $\tau_0$ is the decay time when the satellite that is
initially on an orbit with inclination $\theta=0^{\rm o}$ has lost
90~per cent of its mass, whereas $\tau_{90}$ is the decay time for
polar orbits ($\theta=90^{\rm o}$). }
\label{tab:dectemps}
\end{table}

\subsubsection{Satellites with $M_s=0.2\,M_d$}
The temporal evolution of satellite masses with $M_s=0.2\,M_d$ is
shown in Fig.~\ref{fig:ms1}c. There are no significant differences in
survival times for satellites in spherical and flattened DMHs if
$r_a=55$~kpc.  At the same time, the dependency on the inclination
decreases, causing the range to be narrower in both cases. The cause
is the fast decay of the satellites, so that the anisotropy of the
DMH's velocity dispersion does not have enough time to act.
To better assess this, we introduce a set of computations selecting
larger initial apo-galactic distances (models G3 and G4). The cut-off
radius of the Galaxy is increased, which changes the rotational curve
(see Fig.~\ref{fig:vc}).  The results are also plotted in
Fig.~\ref{fig:ms1}c.  A similar spread of survival times as for models
with $M_s=0.1\,M_d$ and 'G2' flattened DMHs becomes evident; the
range of disruption times for spherical (G3) and flattened DMHs (G4)
are, respectively, $\sim 1$ and $\sim 2$ Gyr. 

The results concerning the disruption times {seen on Fig. 4c between
small and large DMHs (G1/G3 and G2/G4 pairs displayed on Fig.4c, 
bottom panel)} are related to one another as follows.  
 DMHs G3 and G4 have the same mass as
G1 and G2, but are more extended by a factor~$\eta= 133\, {\rm kpc} /84\, {\rm kpc} = 1.58$
(Table~\ref{tab:galmods}).  This implies that the dynamical
time-scale $(\propto 1/\sqrt{G\rho})$, 
i.e. the periods of satellites on equivalent orbits, are
longer in haloes~G3 and~G4 by a factor $\sqrt{\rho(G2)/\rho(G1)} = 1.58^{3/2}=2$. Orbits in G3 and G4 
equivalent to those in G1 and G2, respectively, are orbits with
semi-major axes extended by $\eta$ in a homologous mapping of the systems. 
 Our satellite orbits, however, have
apo-galactic distances in G3 and G4 twice as large as in DMHs G1 and
G2. The orbital times of models G3S2$nn$ and G4S2$nn$ are in total
$1.58^{3/2} \times 2/1.58 = 2\times 2/1.58\approx 2.5$ 
 times longer than models of satellites in DMHs G1
and G2.  This is approximately what we observe from comparing the
curves on Fig. 4c with DMHs G1/G3 or G2/G4. 

On the top panel of Fig.~\ref{fig:ms1}c, 
 the time when $M(t)/M(0) \approx 0.10$ is $t \approx 2$ Gyrs for all 
 G1 models. If the homologous transformation applied strictly, the curves 
for the G2 halo models should approach 5 Gyrs when $M(t)/M(0) = 1/10$. 
The fact that 
they are spread 
between 4 and 5 Gyrs, and thus deviate from the homologous map, 
 indicates that the disc and bulge, which were left unchanged, play an  important  role in the mass decay rate of the satellites.  
 Furthermore, the spread in destruction 
times between models is a factor~$\approx2$, from 1~Gyr (G1 models) to 2~Gyr (G3 models), 
suggesting that the time-scales for orbital decay are
controlled by the DMH, while the combined tidal field of the disc and
bulge contributes mainly to mass stripping. Similar conclusions would apply 
for the G2/G4 models shown on the bottom panel of the figure.  

\subsubsection{Prograde versus retrograde orbits} 
\label{sec:proretro} 
{Results for models with spherical DMHs may be divided into two 
according to whether the orbit of the satellite is aligned with the 
disc's angular momentum vector (prograde) or anti-aligned
(retrograde). Keeping the initial satellite velocity vector unchanged, 
a prograde orbit is found for an initial orbital inclination angle $0^\circ
< \theta < 90^\circ$, and retrograde orbits in the cone $90^\circ <
\theta < 180^\circ$. 

Table 3 lists four models with spherical G1 DMHs and ellipticity $e
= 0.7$ (top segment in the Table). Models G1S100 and G1S1180 are
respectively prograde and retrograde with respect to the disc, but are 
otherwise identical. From Fig. 4a (top panel) we find for these two 
simulations a 90\% mass-loss after $\approx 3$ Gyr and $3.5$ Gyr, 
respectively, an increase of nearly 20\% ; a similar conclusion
applies for models G1S145 and G1S1135. These findings are
qualitatively similar with those of } 
 VW: (i) Satellites on prograde orbits lose angular
momentum faster than their retrograde counterparts, leading to more
rapid decay. (ii) Polar orbits have a similar decay rate as retrograde
orbits,  {as found from comparing model G1S190 and G1S1135, Fig
4a}. This implies that our treatment of the  {live} disc captures the
essential physics relevant for this work. 

 {Figure~\ref{fig:tvstheta} 
summarizes the findings for decay rates for the simulations
performed.} 
Point (i) above also applies  {to} 
 flattened DMHs. However, Fig.~\ref{fig:tvstheta} suggests
{in this case} that  {the difference in decay rates between
prograde and retrograde orbits} 
 is reduced by about 80~\% for flattened DMHs.

For spherical DMHs the above results can be understood partially by
considering Chandrasekhar's expression (Chandrasekhar 1960) for
dynamical friction,
\begin{equation}
{\bf F}_{\rm{df}}=-\frac{4\pi G^2 M_s^2(t) \rho(< v_s) 
 \rm{ln}\Lambda}{\Delta v^3}{\bf v}_s, \label{eqn:dfric}
\end{equation}
$\Delta v=|\vec{v_s}-\vec{v_m}|$ being the relative velocity between
the satellite and the disc particle background, $v_m$ is the disc
particle velocity and $\rho(< v_s)$ the density calculated only for
those particles with velocity less than the
satellite's velocity  $v_s$, and $\rm{ln}\Lambda$ the Coulomb logarithm,
defined as $\Lambda=p_{\rm{max}}/p_{\rm{90}}$. In this expression,
$p_{\rm{max}}$ is the maximum impact parameter (conventionally taken
as the half-mass radius of the system), and $p_{\rm{90}}$ the minimum
impact parameter. Since these quantities are not well defined, the
Coulomb logarithm remains, to a certain degree, an adjustable
parameter. Recent self-consistent computations with different $N$-body
codes leads to ln$\Lambda=1.5 \to 2$ (VW; Fellhauer et al. 2000).

The different decay rate between prograde orbits and their retrograde
counterparts is caused, in part, by the disc's dynamical friction when
the satellite is near perigalacticon.  Retrograde orbits have a much
higher relative velocity $\Delta v$ due to the disc's rotation and,
therefore, they suffer a smaller drag force. The bulge or the DMH's
dynamical friction make no differences since both are non-rotational
and spherical, which also explains the small differences of decay
rates between the polar and the retrograde case (in both cases
dynamical friction through the disc can be neglected compared to the
DMH's dynamical friction). In addition to dynamical friction,
resonances between the satellite and the disc influence the orbital
decay, but a detailed analysis goes beyond the aim of this work.  As
for the different decay rates depending on the satellite's mass, the
dynamical friction force varies with $M_s^2$, so that satellites with
$M_s=0.2\,M_d$ suffer a four times larger friction force than those
with $M_s=0.1\,M_d$.

\subsection{The orbital inclination $\theta$} 
\label{sec:theta}
{Binney (1977) extended the dynamical friction force
(\ref{eqn:dfric}) to non-isotropic velocity fields. He showed how
anisotropic friction leads to orbit alignment with the velocity
ellipsoid plane of symmetry of the host galaxy. Here disc and DMH
spheroids define a unique $z=0$  plane of symmetry, common to both
mass distribution and velocity ellipsoid. We may, therefore,
anticipate enhanced satellite orbit alignment relatively to Binney's
analysis, due to the non-uniform, aspherical mass profile. 

In Fig.~\ref{fig:theta} we graph the time-evolution of the direction angle $\theta$
for a set of simulations with oblate G2 DMHs ($q_h = 0.6$) and
 S1 satellites (solid lines on the figure) as well as two reference
runs with spherical G1 DMHs (dotted lines on the figure).

The average of the orbital inclination $\theta(t)$ decreases monotonically in time
for satellites orbiting in flattened DMHs which have initially
$\theta \ne 0^{\circ}$ or $90^{\circ}$. The decrease in $\theta(t)$ is more
 appreciable for smaller values of $\theta(0)$. This is seen for
instance by comparing the curves with $\theta(0) = 15^{\circ}$ and
$30^{\circ}$ to the solutions with $\theta(0) = 60^{\circ}$ and $90^{\circ}$.
 For the latter, polar orbit,  no decay of $\theta(t)$ is observed for the
duration of the integration, whereas for the $\theta(0) = 15^{\circ}$ case the
orbit aligns fully with the plane of symmetry of the system (coincident
with the disc of the host galaxy).

By contrast, satellites orbiting in spherical DMHs show little or no
 decay of $\theta(t)$, for all initial values of $\theta$ (dotted lines,
Fig.~\ref{fig:theta}). This clearly indicates that the anisotropic DMH, and not
the disc, drives most of the orbital evolution and alignment, since in all cases a
galactic disc is present. 

The figure also reveals periodic oscillations of $\theta(t)$
for satellites on inclined orbits, of frequency approximately in tune
with the satellites' orbital motion. Inspection of the figure shows
this to be the case for systems with either spherical or flattened
DMHs. Note that no such oscillations in $\theta(t)$ is observed for polar
or co-planar orbits. We examine the origin these oscillations, distinguishing
two phases of time around $t = 2$ Gyrs.

For $0 < t < 2$Gyr the satellites orbital radius $r_s \gg R_d$. Over this
 interval of time, the orbits are such that those obtained for flattened
DMHs lead to much larger oscillations in $\theta(t)$ compared with the
solutions with spherical DMHs. We therefore attribute these oscillations to
torques from the DMH acting on the satellites
$$ {\bf \Gamma} = {\bf r} \times {\bf \nabla}\Phi =
 R \left( \frac{\partial\Phi}{\partial z} - z \frac{\partial\Phi}{\partial
R} \right) \, {\bf e}_{\phi} $$
which by symmetry arguments must lie in the plane of the axi-symmetric
galaxy. The torque $\Gamma$ is positive or negative according to the phase
of the orbit.

For $t > 2$Gyr the situation is similar for all calculations,
independently of the morphology of the DMH. Thus the oscillations we observe
clearly for flattened-DMH orbits are now noticeable for the solutions
with spherical DMHs, too. In this phase of evolution, $r_s \sim R_d$
 or less so that the disc potential contributes most of the force
felt by the satellite and hence the torque $\Gamma$ acting on it.
At this stage, a coupling between the disc response and the satellite
motion is expected: we observed that these oscillations are highly
softened in calculations with a static disc and bulge.
 Since the
orbital angular momentum $L \approx  r_s v_s m_s$ and
$\Delta L = \Gamma dt \approx r_s G\Sigma (r_s/v_s)$, where $\Sigma$ is the
disc's surface density, both $L$ and the angular momentum accrued $\Delta
L$ over one revolution will be of comparable magnitude if $v_s^2 \sim
G M_d/r_s$, i.e. when the disc potential is the predominant contributor
to the force acting on the satellite. The direction angle $\theta(t)$
varies therefore wildly towards the end of the simulations in all cases
save the coplanar $\theta(0) = 0^{\circ}$ one,
for which $\Gamma = 0$ at all times. \newline

The oscillations or periodic fluctuations we have discussed are subject to
enhancements owing to our choice of a grid numerical method of
integration. The Cartesian grid code limits the vertical resolution of
a thin disc. Consequently the response of the disc to heating by the
satellite is not correctly quantified. Furthermore, once the remnant
satellite has merged with the disk, the position of it's centre of
density becomes ill-defined by virtue of the satellite ceasing to
exist as a bound entity; $\theta(t)$ will reflect this uncertainty for
$t\simgreat 2-3$~Gyr.  With 32 mesh points spread over a length of $3
R_d$, the position of the centre-of-density and the disc structure are
resolved to $l \simeq 3R_d/32 \sim r_s/10 $ when $r_s \approx R_d$.
Hence the error on the angle $\theta$ may be estimated to be
$\sin\theta \approx \theta = l/r_s \sim 1/10$ or $5^{\circ}$
approximately. This puts into perspective the magnitude of the oscillations
seen on Fig.~6 for $t \simgreat 2-3$~Gyr, though without accounting
for them fully. This leads us to conclude that the physical effect of
the torque $\Gamma$ by the disc on the satellite is qualitatively
correct, although the quantities somewhat uncertain.

\begin{figure}
\vspace{9.0cm}
\includegraphics{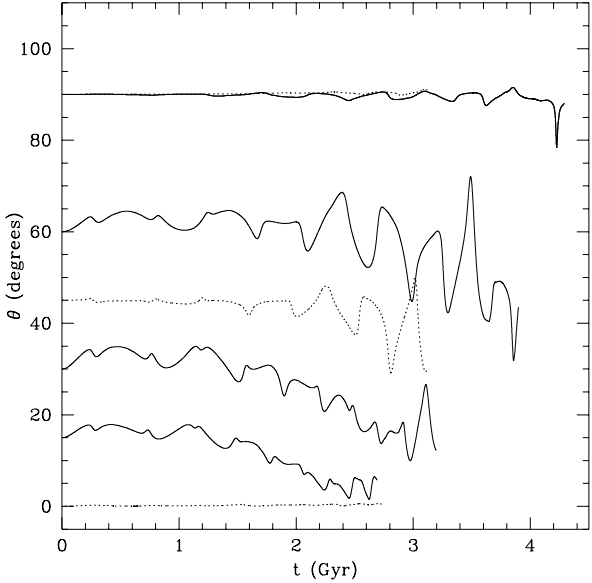}
\caption{Evolution of the orbital inclination for models G2S100,
G2S115, G2S130, G2S190 (full lines, satellites within the flattened
DMH) and G1S145, G1S190 (dotted lines, satellites within the
spherical DMH), until they retain 10\% of their initial mass. } 
\label{fig:theta}
\end{figure}

\subsection{Orbital eccentricity}
\label{sec:orbecc}
In Fig.~\ref{fig:orbecc} we plot the eccentricity evolution for
satellites with mass $M_s=0.1\,M_d$. The eccentricity is calculated by
fitting straight lines to $r_a(t)$ and $r_p(t)$
(e.g. Fig.~\ref{fig:orbecc}a) and interpolating $e(t)$ until the
satellite has $10~\%$ of its initial mass.

The orbital eccentricity does not remain constant as dynamical
friction shrinks the orbit.  The evolution of $e(t)$ depends on
$e(t=0)$ and $\theta (t=0)$, but from Fig.~\ref{fig:orbecc}b we
observe that the general behaviour is for the orbits to become more
radial with time.  The only clearly evident exception is prograde
model G2S100 ($e(0)=0.7$), which shows a pronounced decrease of
$e(t)$. In this case, dynamical friction from the flattened DMH plus
disc is so large that the apo-galactic distance decreases much faster
than the peri-galactic distance, as we can observe in
Fig.~\ref{fig:orbecc}a. Close inspection shows that this is merely the
extreme of a general trend. Comparing the co-planar prograde orbits
($\theta=0^{\rm o}$: GnS100, GnS100e, GnS100c; n$=1,2$) with the polar
orbits ($\theta=90^{\rm o}$: GnS190, GnS190e, GnS190c), it is evident
that the former show a stronger sensitivity on initial eccentricity than the
latter. The effect is such that circular co-planar prograde orbits
gain eccentricity rapidly, whereas eccentric ($e(0) \approx 0.7$) prograde co-planar orbits
circularize. Disc--satellite coupling via dynamical friction and
induction of spiral modes in the disc and associated transfer of
angular momentum between satellite and disc are the likely reason, but
we do not dwell longer on this, as disc-satellite coupling is not the
main topic of this work, which in any case does not resolve the disc
vertical structure. We merely state here that the data in
Fig.~\ref{fig:orbecc}b suggest that there is a stable eccentricity,
$e_{\rm stab}\approx0.5$ for co-planar prograde orbits in our
flattened DMH, such that $e(t)$ increases when $e(0)<e_{\rm stab}$,
whereas $e(t)$ slighty decreases when $e(0) > e_{\rm stab}$, and $e(t)\approx
e_{\rm stab}$ for all $t$ until satellite disruption,  if
$e(0)\approx e_{\rm stab}$.  However, if $e(0)$ is smaller than $e_{\rm
stab}$, $e(t)$ does not remain close to $e_{\rm stab}$ once it has
reached this critical eccentricity. For our spherical DMH, $e_{\rm
stab}\approx0.6$. 

This behaviour agrees with that found by van den Bosch et al (1999). They perform numerical calculations 
 using a galaxy models similar to G1, with satellite masses on the order of that of our models S1. They 
observe that the eccentricity remains remarkably constant for $e(0)\geq 0.6$. 
Unfortunately, they do not include calculations with lower initial eccentricity that we can compare with. We note in passing that Prugniel \& Combes (1992) already found that
initially circular orbits rapidly acquire eccentricity, as is also 
evident in the calculations of Fellhauer et al. (2000). 

\begin{figure}
\vspace{9.0cm} \includegraphics{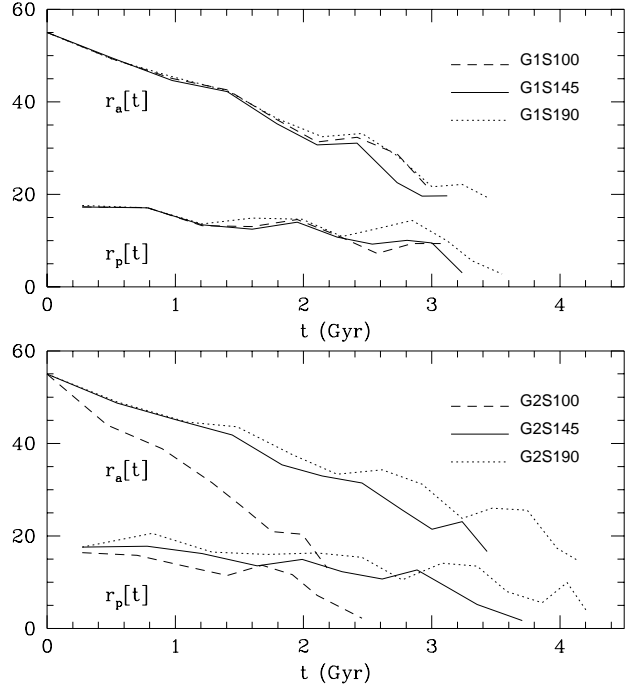} 
\caption{{\bf a:} Evolution of the apo-galacticon and perigalacticon
distance (in kpc) for the models with $e(t=0)=0.7$ until the satellite
has $10~\%$ of its initial mass. }
\label{fig:orbecc}
\end{figure}

\begin{figure}
\vspace{8.0 cm} \includegraphics{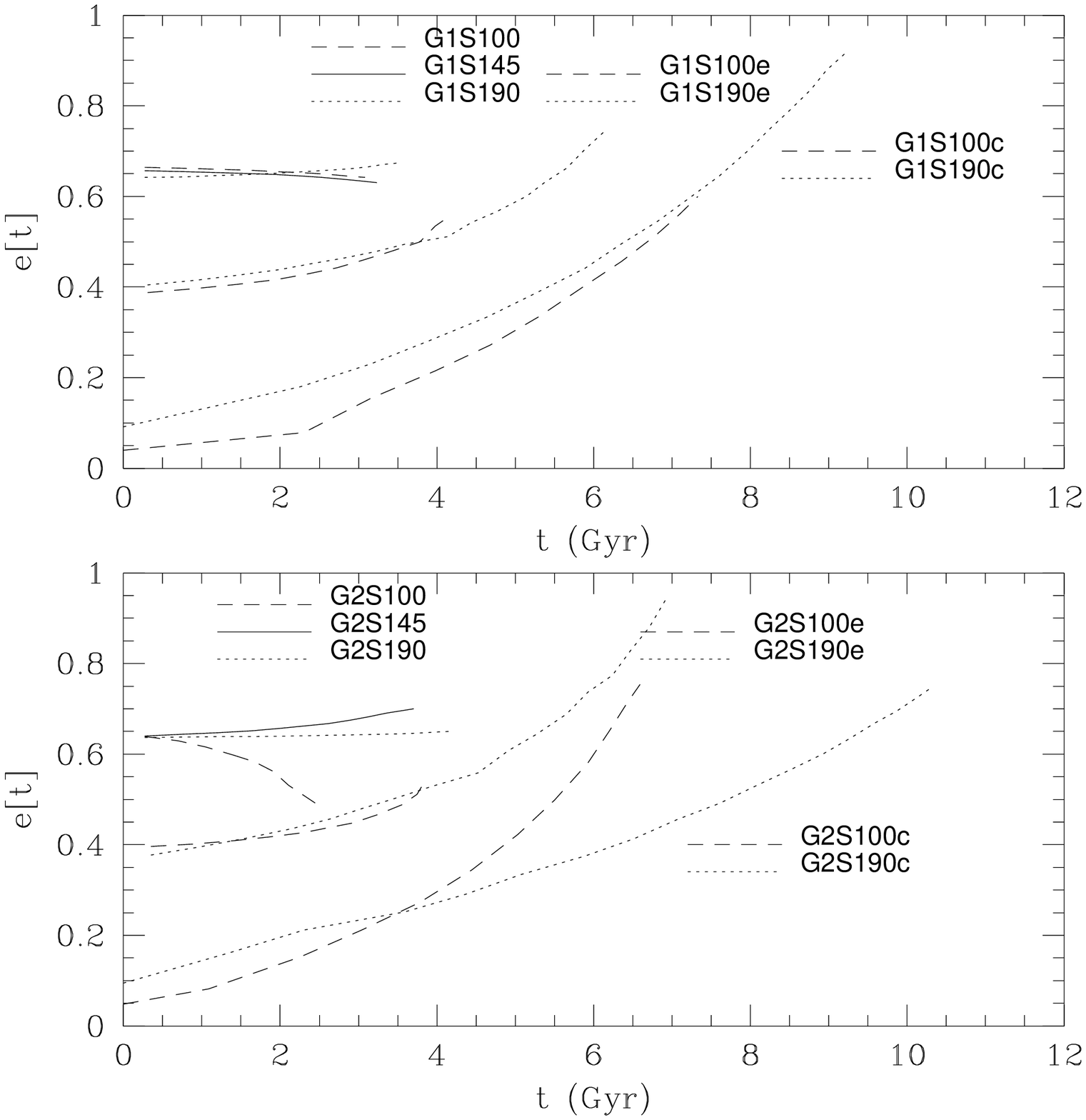} \contcaption{ {\bf b:} The eccentricity
evolution for the models shown in
Fig.~\ref{fig:orbecc}a. $e(t)=1-r_p(t)/r_a(t)$ and is measured by
fitting the curves $r_a(t)$ and $r_p(t)$ by straight lines. We also
plot it for the models with $e(0)=0.45$ and $e(0)=0$ (circular
orbits), calculated following the same scheme.  }
\end{figure}
 
\subsection{Orbital precession}    
\label{sec:orbprec}
The orbital plane of a satellite and its unbound particles precesses
in a flattened potential which smears out the tidal debris stream.  The precession angle, $P(t)$,
is calculated by projecting the orbital angular momentum vector onto
the galactic xy plane and measuring its change with time.  In
Fig.~\ref{fig:prec} we plot $P$ for some of our models. The
precession, $dP/dt$, increases at later times due to the anisotropy of
the disc's potential, the satellite having decayed to its vicinity.

As expected, flattened DMHs lead to larger precession. Comparing
models G1S145 (satellite within a spherical DMH) and G2S145
(satellite within a flattened DMH), we observe that the change of $P$
is, respectively, $\simeq 50^\circ$ and $\simeq 150^\circ$, i.e,
approximately three times larger at $t=3$ Gyr. Since the DMH is
spherical for models with G1 the precession of the orbital plane is
due to the disc gravitational quadrupole moment.  The orbital plane
precesses faster the smaller its inclination is, orbits with
$\theta\simless 45^o$ precessing by 180$^o$ in 3~Gyr.  Polar orbits do
not precess at all.

\begin{figure}
\vspace{9 cm}
\includegraphics{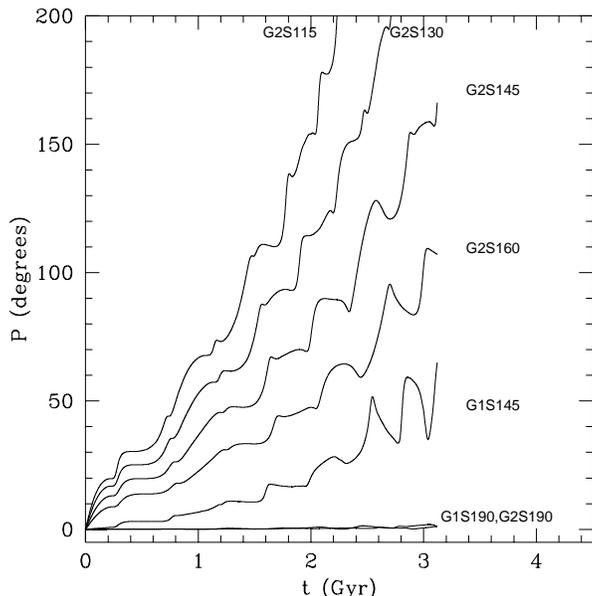}
\caption{The precession angle $P$ for some of our models. } 
\label{fig:prec}
\end{figure}

\subsection{Tidal streams}
\label{sec:str}
The accretion history of the Milky Way and other major galaxies leaves
signatures in the form of old tidal streams in the DMHs of these
galaxies as found in observational surveys such as that of Dohm-Palmer et al (2001), or Mart\'inez-Delgado et al (2001). The detection of the Sagittarius dwarf tails (Iabata et al 1994) therefore likely is a generic features of large galaxies. 

Theoretical models of this process have shown good agreement with observations (Helmi \& White 1999; Zhao et al. 1999; Helmi \& de Zeeuw
2000). The changes in orbital inclination $\theta$ and the orbital
precession in flattened systems imply that the tidal debris emanating
from a disrupting satellite will significantly spread out in $\theta$,
which will make reconstruction of the accretion history of a major
galaxy difficult if its DMH is flattened.

In Fig.~\ref{fig:debris1} we plot the deviation angle of the
satellite's particles from the initial orbital plane in three time
snaps. This is done for models G1S145 and G2S145
(Fig.~\ref{fig:debris1}a, $\theta(0)=45^\circ$), and for G1S190 and
G2S190 (Fig.~\ref{fig:debris1}b, $\theta(0)=90^\circ$). The first
time-snap shows satellite particles after first passage through
perigalacticon at $t=0.62$~Gyr, the second one is at an intermediate
time ($t=1.52$ Gyr) while the last frame is at a late stage of the satellite
orbit.  The debris does not remain in the initial orbital
plane. This effect becomes more pronounced the closer the satellite is
to the galaxy's centre, when the mass loss (Fig.~\ref{fig:ms1}) and
the oscillations of the orbital inclination (Fig.~\ref{fig:theta})
primarily occur, and the larger the number of perigalacticon passages
is.  From Fig.~\ref{fig:debris1}{\bf a} we also observe that the deviations from the orbital plane are
enhanced when the DMH is flattened since satellite 
orbits within
oblate DMHs align with the symmetry plane (i.e. $\theta(t)
\rightarrow 0$).  Fig.~\ref{fig:debris1}{\bf b} shows that the spread of
satellite debris is much smaller for satellites in polar orbits than
for those with intermediate inclinations, since inclination decay and
oscillations vanish for polar orbits.

\begin{figure}
\vspace{8.5cm}
\includegraphics{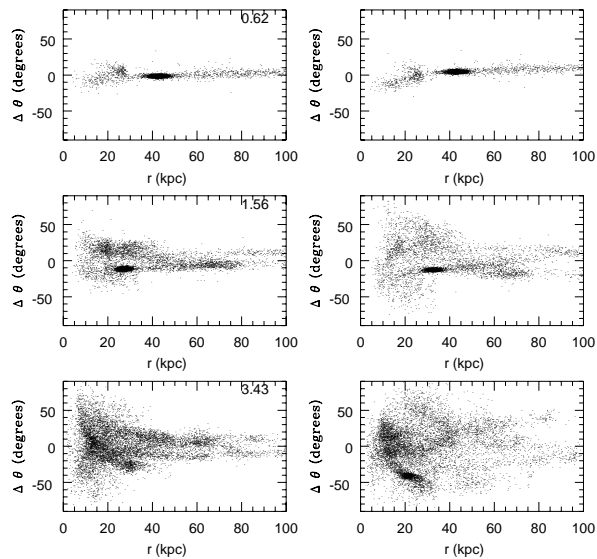}
\caption{{\bf a:} Deviation angles for all satellite particles from
the initial orbital plane ($\theta=45^\circ$). The left column depicts
model G1S145 (spherical DMH), and the right column shows G2S145 (flattened
DMH). Rows show three time snaps (given in Gyr). In the last one,
the satellite has been fully destroyed. }
\label{fig:debris1}
\end{figure}  

\begin{figure}
\vspace{8.5cm} \includegraphics{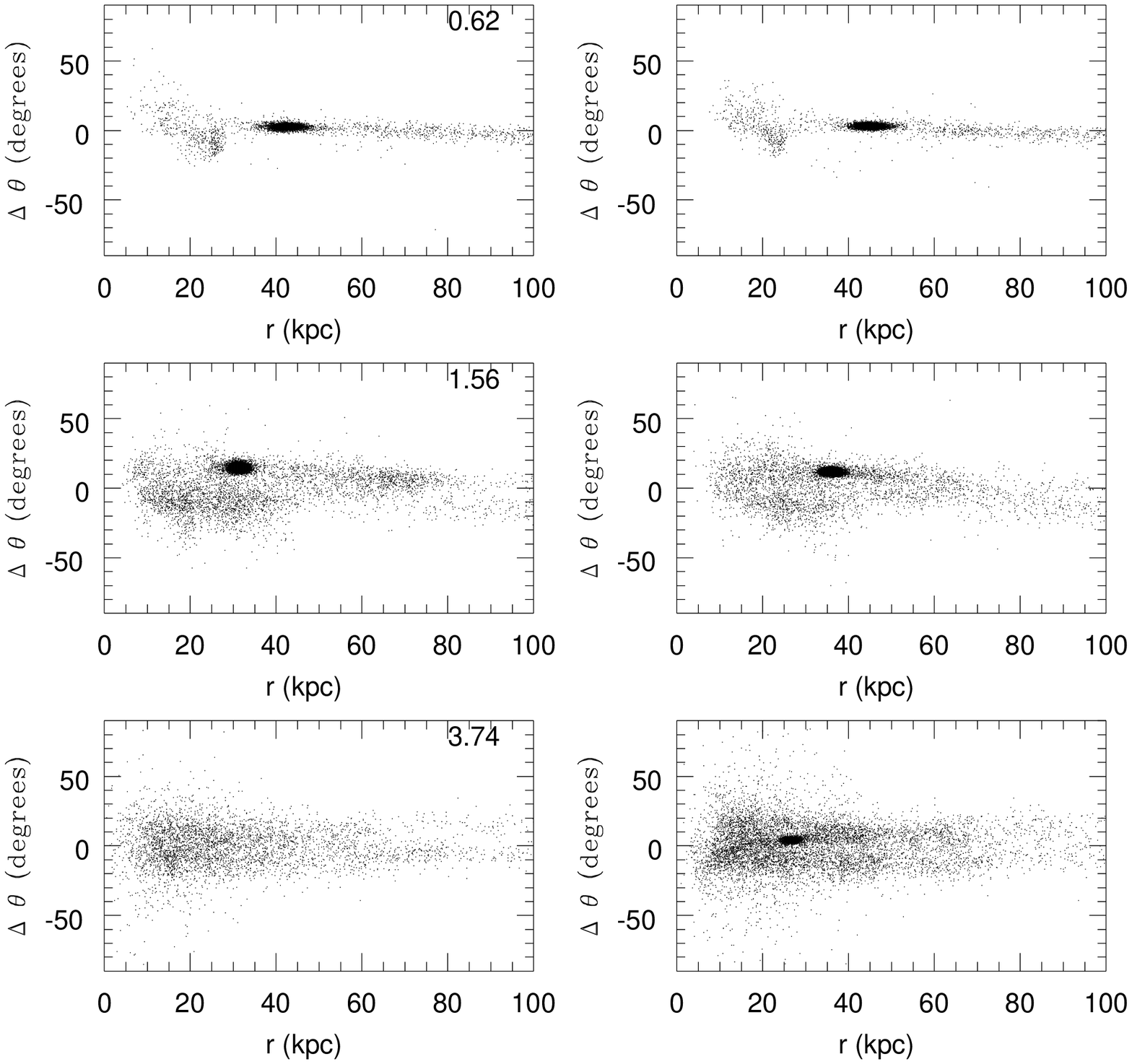} \contcaption{{\bf b:} As
Fig.~\ref{fig:debris1}a but for models G1S190 (spherical DMH) and
G2S190 (flattened DMH), with initial inclination $\theta=90^\circ$.}
\end{figure}

\section{CONCLUSIONS}
In order to assess the importance of dynamical friction in extended 
oblate DMHs on the distribution of satellite galaxies around their
primary, we perform self-consistent $N$-body computations of
satellite galaxies with masses amounting from 10~to 20~per cent of the
primary's disc. The satellites are placed on different orbits
in spherical and flattened DMHs that have embedded galactic discs and
bulges.

The calculations with spherical DMHs lead to results in good
        agreement with
        those obtained by Velazquez \& White (1999). Modest differences
        in quantities are attibuted to the increased mass resolution
        of our calculations compared with theirs, as well as
        different  linear resolution (grid size versus smoothing length of their
        TREE algorithm).

        Satellites evolving in spherical DMHs on prograde orbits relatively
        to the primary galaxy's disc rotation decay faster than satellites
        on retrograde orbits or on polar orbits. This results from orbital
        resonances between the disc and the satellites.

Of particular interest, however, is that our results demonstrate that
non-isotropic dynamical friction in flattened DMHs works as a removal
mechanism of satellites with low-inclination orbits, whereas it
enhances the survival time of satellites on near polar orbits. Thus,
satellites on polar orbits survive about 70~per cent longer than
satellites on orbits that have a small inclination relative to the
primary galaxy's disc (Table~\ref{tab:dectemps}), irrespective of the
relative orbital sense (Fig.~\ref{fig:tvstheta}) in an oblate DMH with
axis ratio $q_h=0.6$. This is the key result of this investigation. 

        This result helps understand the distribution of dwarfs
        galaxies in the Milky Way. Since they are mainly distributed
        near the galactic pole (Carney et al. 1987) we may infer a
        selection of survivor dwarfs from a primordial population.
        The accelerated orbital decay and alignment with the disc
        of dwarfs within a flattened halo would go some way towards
        accounting for the data. However if the masses deduced for
        these satellites (~ $10^8$ solar, compared with $10^9$ for our
        models) is a good measure of their mass
        at the formation time, our computations indicate times as long
        as a Hubble time for effective mergers. Discrepancies in
        timescale may well be accounted for if we substitute for the
        isothermal halo the more concentrated  NFW 
(Navarro, Frenk \& White 1995) 
models or haloes with a steeper cusp (Moore et al. 1998): when each
        halo model is scaled to the same integrated mass inside the
        solar radius,
        the particle
        velocity dispersion in these models
        drops faster with radius than for isothermal
        spheres. Because of the strong dependence of friction on
        velocity dispersion, this would reduce the timescale for orbital decay
        very much and offset the effect of reduced satellite masses. We have
        not, however, performed calculations with different halo mass profiles.

Our computations further show that satellites on orbits with
eccentricity $e\simgreat 0.5$ and with masses larger than 10~per cent
of their primary galaxy's disc merge within only a few~Gyr with the
primary galaxy.  The time it takes to merge increases with decreasing
orbital eccentricity (Fig.~\ref{fig:tvstheta}). We therefore deduce
that massive satellites around distant galaxies, such as typically enter
the samples that show the Holmberg effect, may be preferentially on
near-circular polar orbits or on orbits with apo-galactica further
away from their primary galaxy than about 130~kpc.  

The calculations also suggest that there exists a critical
eccentricity, $e_{\rm stab}$, for co-planar prograde orbits such that
initial eccentricities that are close to $e_{\rm stab}$ remain within
the vicinity of this 'stable' value, whereas initial orbital
eccentricities that differ from $e_{\rm stab}$ evolve towards the
critical value but the orbit keeps evolving past $e_{\rm stab}$
(Section~\ref{sec:orbecc}).  For our oblate DMH, $e_{\rm
stab}\approx0.5$, whereas for the spherical DMH $e_{\rm
stab}\approx0.6$.

We also note that the high precession rates of satellite
orbits in flattened DMHs and the decrease in orbital inclination
leads to tidal debris streams being completely smeared apart for
initially inclined orbits.
 
 We want to comment that, despite our use of only two values for the satellite mass in our calculations, this range seems to be representative to reproduce the typical mass of the satellite that Holmberg (1969) and Zaritsky \& Gonz\'alez (1999) find in their observations when the initial apo-galactic distances is selected properly (Ibata et.al 2001). As Tormen (1997) finds in his numerical calculations of hierarchical galaxy clusters history, more massive satellites ($\sim 10^{11}$) are unlikely to survive due to the large drag force they suffer. On the other hand, though less massive satellites ($\sim 10^8$ solar) feel a negligible drag force, they are quickly disrupted after some peri-galacticon passages due to their low binding energy.

        This paper has sought to quantify the effect of aspherical DMHs
        on the orbits of galactic satellites. The analysis suggests
        enhanced Holmberg decay, yet what can we say of a population of
        satellites as a whole?
       Our model satellites require a few  orbits
 around the host galaxy if dynamical friction is to be effective.
Thus within  one Hubble time a satellite would require
= 5 revolutions (say) or $t$ =
2 Gyr for a single revolution at most. In the Milky Way
the orbital time  $t = 200$ Myr at $r =$
10kpc; assuming an isothermal halo with  $\rho\propto  r^{-2}$,
the critical orbital time $t$ = 2 Gyr would be found at $r$ =
50 kpc or so. In other words, satellites that are too far from the host
galaxy will not have time to experience dynamical friction and hence will
not have suffered Holmberg decay. On the other hand, satellites closer to
their host galaxy will merge quickly through the process described here.
 Zaritsky et al. (1999) have noted that satellite populations tend to
remain isotropically distributed for satellites with $r >$ 50 Kpc.

A more elaborate study is under way, and ultimately we aim at making a
statistical study of a modelled observational sample to infer if the
Holmberg effect can indeed be produced by flattened DMHs.

\section{Acknowledgements} 
We thank Francine Leeuwin and Holger Baumgardt 
for  useful comments and Mike Fellhauer
for his help with {\sc Superbox}. We are also grateful to H\'ector
Vel\'azquez for providing helpful information on his work and to the anonymous referee for his useful comments. JP and CMB
acknowledge support through an SFB~439 grant at the University of
Heidelberg.

\end{document}